\documentclass[preprint,aps,showpacs,showkeys,amsmath,amssymb,nofootinbib]{revtex4}

\usepackage{graphicx}%
\usepackage{array}%
\usepackage{bm}%

\usepackage{latexsym}%

\begin{document}


\title{A new molecular dynamics calculation and its application to the spectra of light and strange baryons}

\author{Takashi Watanabe}
 \email{watanabe@ph.noda.tus.ac.jp}
\author{Makoto Oosawa}
\author{Koichi Saito}
 \email{ksaito@ph.noda.tus.ac.jp}
\author{Shinsho Oryu}
 \email{oryu@ph.noda.tus.ac.jp}

\affiliation{
Department of Physics, Faculty of Science and Technology, 
Tokyo University of Science, Noda, Chiba 278-8510, Japan}

\date{\today}

\begin{abstract}
A new approach based on antisymmetrized molecular dynamics is proposed to correctly take account of the many-body correlation. 
We applied it to the spectra of low-lying, light and strange baryons.  The inclusion of the quark-quark correlation 
is vital to predict the precise spectra, and the semi-relativistic kinematics is also important to correct the level ordering. 
The baryon spectra calculated by the present method is as precise as the Faddeev calculation. 
\end{abstract}

\maketitle


Although recent lattice calculations based on Quantum Chromodynamics (QCD) start providing reliable hadron spectra, it is still difficult 
to obtain precise, detailed predictions of the physical states~\cite{lattice}.  
Thus, QCD-inspired, effective models are still useful to get some insight into many 
phenomena of hadronic systems. The central issue to be addressed is then the quantitative description of 
low-energy phenomena, e.g., hadron spectra, baryon-baryon interactions, in-medium change of hadron properties~\cite{saito}, etc. 

Among QCD-inspired models, the simplest approach is the nonrelativistic (NR), constituent quark model (CQM), and it is 
doubtlessly successful in describing the ground-state properties of hadrons and baryon-baryon interactions at low energies~\cite{cqm}. 
There are, however, some obscure problems such as the level-ordering problem in excited hadron spectra, etc. 
In addition to ordinary mesons and baryons, the (NR) CQM is also applied to exotic states, e.g., multiquark states 
including pentaquarks, non $q{\bar q}$-mesons (tetraquarks~\cite{tetra}), etc. 

The QCD lagrangian for massless up and down quarks is chirally symmetric, and the axial symmetry is spontaneously broken
as evidenced in the absence of parity doublets in the low-mass hadron spectra. This implies the existence 
of the massless Nambu-Goldstone (NG) bosons, e.g., the pions. The non-zero pion mass is then a consequence of
the fact that the light quark has a small mass, which gives the explicit symmetry breaking.
Thus, one arrives at a low-energy scenario that consists of the NG bosons and constituent quarks (with a rather massive, 
constituent quark mass in the NG phase) interacting via a force governed by spontaneously broken, approximate chiral symmetry. 
Thus, it is of very importance to take into account the effect of the NG-boson exchange as well as the usual one-gluon-exchange 
(OGE) force in the CQM. 
Note that multigluon degrees of freedom can be elminated by introducing a confining potential. Furthermore, the relativistic effect 
may be vital to produce the hadron spectra, because the quarks move in the region of the hadron size. 

There are a lot of model calculations to study hadron spectra, baryon-baryon interactions, etc~\cite{cqm,SVM1,ESP1,ESP3,ESP4}. 
In the calculation of baryon spectra, the most precise method may be based on the Faddeev equations for the three constituent quarks~\cite{ESP3}. 
However, for multiquark states like the exotic hadrons, it is quite difficult to perform such 
precise calculations.  Thus, it is very important and useful to construct a powerful method which allows us to precisely 
calculate the wavefunction even for multiquark states. 

Molecular dynamics (MD) is very successful in treating many-body systems. In particular, fermionic molecular dynamics 
(FMD) was first developed by Feldmeier~\cite{fmd} to describe the ground states of atomic nuclei and heavy ion reactions in 
the energy regime below particle production. 
Antisymmetrized molecular dynamics (AMD)~\cite{ONO,ENYO1} is very similar to FMD with respect to the choice of the trial 
state. In the past decade, AMD has been applied to various studies of 
light or medium nuclei~\cite{ENYO21}. In AMD, it is not necessary to take any model assumption (like shell, cluster, etc.), and 
the AMD wavefunction can simultaneously describe a variety of nuclear structure by following the variational principle.  

However, AMD has not yet been successfully applied to a few-body system, 
because the trial wavefunction in AMD is given by the Slater determinant of single-particle, gaussian wave packets and hence 
the correlation among nucleons is missing.  To improve this weak point, some of the present authors have recently proposed a modified version of AMD, in which 
the Jacobi coordinates and the generator-coordinate method (GCM) are introduced to describe the correlation~\cite{JAMD}.\footnote{
The nucleon-nucleon correlations are also considered in the stochastic variational method (SVM)~\cite{SVM1,SVM2} or 
the coupled-rearrangement-channel Gaussian-basis variational method (CRCGV)~\cite{CRCGV1}.} 
We here refer to this method as the Jacobi-coordinate-basis AMD (JAMD). 
In JAMD, one can easily treat the many-body correlations.  Furthermore, it is possible to extract the center-of-mass (c.m.) 
wavefunction and remove the zero-point energy. 

In this paper, we apply the present method to the SU(3) octet- and decuplet-baryon spectra, 
and compare the JAMD result with that of 
the simple AMD or the Faddeev result to illustrate how this method is useful.


In AMD, the wavefunction of $A$-quark system with definite ($\pm$) parity, $|\Psi^\pm \rangle$, is given by~\cite{ONO,ENYO1}
\begin{equation}
|\Psi^{\pm}\rangle = 
|{\mathcal C}\rangle \sum_{i=1}^{A!}|\Phi_{p_i}^{\pm}\rangle|{\mathcal S}_{p_i}\rangle|{\mathcal F}_{p_i}\rangle , 
\label{wf1}
\end{equation}
where $|\Phi_{p_i}^\pm \rangle$, $|{\mathcal S}_{p_i}\rangle$, $|{\mathcal F}_{p_i}\rangle$ 
and $|{\mathcal C}\rangle$ are, respectively, the spatial, spin, flavor and color wavefunctions. 
The color wavefunction is then antisymmetrized as
\begin{equation}
|{\mathcal C}\rangle =\sum_{i=1}^{A!}{\rm sgn}(p_i)
|m^{c}_{p_i(1)}m^{c}_{p_i(2)}\cdots m^{c}_{p_i(A)}\rangle ,  \label{cl1} 
\end{equation}
where $m^c$ specifies the color and 
sgn$(p_i)$ is the sign of the permutation $p_i \, (i = 1, 2, \, \cdots, \, A!)$:
\begin{eqnarray}
p_i=
\left(\begin{array}{ccccc}
1&2&3&\dots&A \\
j_1&j_2&j_3&\dots&j_A
\end{array}\right)
\equiv
\left(\begin{array}{ccccc}
1&2&3&\dots&A \\
p_i(1)&p_i(2)&p_i(3)&\dots&p_i(A)
\end{array}\right)  . \label{perm}
\end{eqnarray}
Because the color wavefunction is already antisymmetrized, the other wavefunctions must be symmetrized in Eq.(\ref{wf1}). 
The spatial wavefunction is then given by the product of a single-particle wavefunction 
\begin{eqnarray}
|\Phi_{p_i}^{\pm}\rangle = (1\pm P)|\phi_{1}(p_i(1))\phi_{2}(p_i(2))\dots \phi_{A}(p_i(A)) \rangle, \label{swf1}
\end{eqnarray}
where $1\pm P$ is the parity projection operator and 
the $i$-th single-particle wavefunction (at ${\vec r}_i$) is given by a gaussian function 
\begin{eqnarray}
\phi_i(j)=\exp[-\nu({\vec r}_i-{\vec Z}_j )^2], 
\end{eqnarray}
with ${\vec Z}$ the center of the wave packet and $\nu$ a variational parameter for its width. 
The spin wavefunction is expressed as
\begin{equation}
|{\mathcal S}_{p_i}\rangle = 
\sum_{m^s = \uparrow , \downarrow} c^s_{p_i(1) \cdots p_i(A)}
|m^s_{p_i(1)} \cdots m^s_{p_i(A)}\rangle , \label{spin1} 
%
\end{equation}
where the coefficient, $c^{s}$, is a variational parameter.  However, since, for the low-lying baryon states, the spin-flavor 
structure ($|{\mathcal S}_{p_i}\rangle |{\mathcal F}_{p_i}\rangle$ in Eq.(\ref{wf1})) can be given by SU(6) symmetry, 
we here use it to reduce the computation time. 

In JAMD, the spatial wavefunction is expressed in terms of gaussian functions with the Jacobi coordinates 
(${\vec \rho}_i; \, i= 1, 2, \, \cdots, \, A$)~\cite{JAMD}.  
Then, the usual single-particle coordinates, ${\vec r}_i$, can be related to the 
Jacobi coordinates through ${\vec \rho}_i=\sum_{k=1}^A K_{ik}{\vec r}_k$, where the matrix $K$ is given by~\cite{SVM2}
\begin{eqnarray}
K=
\left(\begin{array}{ccccc}
1&-1&0&\cdots&0 \\
\frac{m_1}{m_{12}}&\frac{m_2}{m_{12}}&-1&\cdots&0 \\
\vdots&&&\vdots\\
\frac{m_1}{m_{12\cdots A-1}}&\frac{m_2}{m_{12\cdots A-1}}&\cdots&\cdots&-1 \\
\frac{m_1}{m_{12\cdots A}}&\frac{m_2}{m_{12\cdots A}}&\cdots&\cdots&
\frac{m_A}{m_{12\cdots A}} \\
\end{array}\right) , \label{kmatrix}
\end{eqnarray}
%
%
where $m_{12\cdots k} = m_1 + m_2 + \cdots + m_k$ and $m_i$ is the mass of the $i$-th particle. 
The center of the gaussian wave packet, ${\vec X}_i$, in the JAMD wavefunction is also related to ${\vec Z}_j$ using the $K$ 
matrix~\cite{JAMD}.  
Thus, the spacial part of the JAMD wavefunction, $\Phi^J$, with the width, $\nu_{ij}^{J}$, can be expressed in terms of 
the usual coordinates as 
\begin{eqnarray}
\Phi^J &=& \exp\Big[-\sum_{i,j=1}^A\nu_{ij}^{J}
({\vec \rho}_i-{\vec X}_i)\cdot({\rho}_j-{\vec X}_j)\Big] 
= \exp\Big[-\sum_{i,j=1}^A\nu_{ij} ({\vec r}_i-{\vec Z}_i)\cdot({\vec r}_j-{\vec Z}_j)\Big] \nonumber\\ 
&=& \exp\Big[-\frac{1}{2}\sum_{i,j=1}^A\tilde{\nu}_{ij}
\Big\{({\vec r}_i-{\vec Z}_i)-({\vec r}_j-{\vec Z}_j)\Big\}^2 -\nu_{AA}({\vec \rho_A}-{\vec X}_A)^{\,2}\Big] , \label{jswf0} 
\end{eqnarray}
where each element of the width matrix, ${\tilde \nu}_{ij}$, is treated as a variational parameter, and 
$\nu^J_{ij}$, $\nu_{ij}$ and ${\tilde \nu}_{ij} \ (i,j=1,2, \, \cdots, \, A)$ satisfy the following relation: 
\begin{equation}
\nu_{ij} = \sum_{k,l=1}^A
\nu_{kl}^{J}K_{ki}K_{lj} = 
\sum_{k=1}^A \tilde{\nu}_{ik}\delta_{ij}-\tilde{\nu}_{ij}
+\frac{m_im_j}{M^2}\nu_{AA} . \label{relation1}
\end{equation}
%

%
%

Now let us apply the JAMD method to the SU(3) octet and decuplet baryons.  The hamiltonian is given by 
\begin{eqnarray}
H=T+V_{\chi}(\vec{r}_{ij})+V_{\rm OGE}(\vec{r}_{ij})+V_{\rm CON}(\vec{r}_{ij}), \label{hamiltonian1}
\end{eqnarray}
where $T$ is the kinetic-energy term.  In this paper, as well as the NR form, we consider the semi-relativistic (SR) form 
to take into account the relativistic effect: 
\begin{eqnarray}
T=\left\{
\begin{array}{l>{$}l<{$}}
\frac{1}{2}\sum_{i>j}^A\frac{m_i+m_j}{m_im_j}({\vec p}_j-{\vec p}_i)^2 
+\sum_{i=1}^Am_i
& \ \ \ for NR, \\
\sum_{i=1}^A\sqrt{{\vec p}_i^{\ 2}+m_i^2}
& \ \ \ for SR, 
\end{array}
\right.  \label{T1}
\end{eqnarray}
where $m_i$ is the $i$-th quark mass.  

The potential generated by the exchanges of the NG bosons ($\pi, K, \eta$) and the $\sigma$ meson is 
given by $V_{\chi}({\vec r}_{ij})=V_{\pi}({\vec r}_{ij})
+V_{\sigma}({\vec r}_{ij})+V_{K}({\vec r}_{ij})+V_{\eta}({\vec r}_{ij})$~\cite{ESP1,ESP3},
where\footnote{It is possible to treat the LS or tensor force in the JAMD method~\cite{watanabe}. 
}
\begin{eqnarray}
V_{\pi}({\vec r}_{ij}) &=& 
\frac{g_{ch}^2}{4\pi}\frac{m_{\pi}^3}{12m_i m_j}
\frac{\Lambda_{\pi}^2}{\Lambda_{\pi}^2-m_{\pi}^2} 
\Big[Y(m_{\pi} r_{ij})-\frac{\Lambda_{\pi}^3}{m_{\pi}^3}
Y(\Lambda_{\pi} r_{ij})\Big]
({\vec \sigma}_i \cdot {\vec \sigma}_j)
\sum_{a=1}^{3}(\lambda_i^a \cdot \lambda_j^a) , \label{pion1} \\
V_{\sigma}({\vec r}_{ij}) &=& 
-\frac{g_{ch}^2}{4\pi}
\frac{\Lambda_{\sigma}^2}{\Lambda_{\sigma}^2-m_{\sigma}^2} m_{\sigma}
{\Big[Y(m_{\sigma} r_{ij})-\frac{\Lambda_{\sigma}}{m_{\sigma}}
Y(\Lambda_{\sigma} r_{ij})}{\Big]}, \label{sigma1} \\
V_{K}({\vec r}_{ij}) &=& 
\frac{g_{ch}^2}{4\pi}\frac{m_{K}^3}{12m_i m_j}
\frac{\Lambda_{K}^2}{\Lambda_{K}^2-m_{K}^2} 
{\Big[Y(m_{K} r_{ij})-\frac{\Lambda_{K}^3}{m_{K}^3}
Y(\Lambda_{K} r_{ij})}{\Big]}
({\vec \sigma}_i \cdot {\vec \sigma}_j)
\sum_{a=4}^{7}(\lambda_i^a \cdot \lambda_j^a) , \label{K1} \\
V_{\eta}({\vec r}_{ij}) &=& 
\frac{g_{ch}^2}{4\pi}\frac{m_{\eta}^3}{12m_i m_j}
\frac{\Lambda_{\eta}^2}{\Lambda_{\eta}^2-m_{\eta}^2} 
{\Big[Y(m_{\eta} r_{ij})-\frac{\Lambda_{\eta}^3}{m_{\eta}^3}
Y(\Lambda_{\eta} r_{ij})}{\Big]} ({\vec \sigma}_i \cdot {\vec \sigma}_j) \Theta_P , \label{eta1}
\end{eqnarray}
with ${\vec r}_{ij}={\vec r}_{i} - {\vec r}_{j}$, $\lambda^a$ the SU(3) generator, $g_{ch}$ the chiral coupling constant, 
$\Lambda_{\pi, \sigma, K, \eta}$ the cutoff parameter, $Y(mr)$ the Yukawa function and 
$\Theta_P (= \cos\theta_P(\lambda_i^8 \cdot \lambda_j^8)-\sin\theta_P)$ 
the mixing term for considering the physical $\eta$.
The masses of the NG bosons are denoted by $m_\pi$, $m_K$ and $m_\eta$. The $\sigma$ mass is chosen to be 
$m_\sigma^2 \sim m_\pi^2 + 4m_{u,d}^2$~\cite{ESP3}. 

The potential due to the OGE is~\cite{ESP1,ESP3}  
\begin{eqnarray}
V_{\rm OGE}({\vec r}_{ij}) &=& 
\frac{1}{4}\, \alpha_s
{\vec \lambda}_i\cdot{\vec \lambda}_j \frac{1}{r_{ij}}
 +\frac{1}{4}\,\alpha_s
{\vec \lambda}_i\cdot{\vec \lambda}_j
\frac{1}{m_i m_j}\frac{e^{-r_{ij}/r_0}}{r_{ij}r_0^2} \nonumber \\
&&
\times
 \left\{
\begin{array}{l>{$}l<{$}}
\frac{1}{6}{\vec \sigma}_i \cdot {\vec \sigma}_j
& \ \ for NR, \\
\frac{1}{4}\Big(1+\frac{2}{3}{\vec \sigma}_i 
\cdot {\vec \sigma}_j\Big)
& \ \ for SR , 
\end{array}
\right.  \label{oge1}
\end{eqnarray}
with $\alpha_s$ the quark-gluon coupling constant and $r_0$ a parameter to be fixed from the data. 
Note that in Ref.~\cite{ESP3} $r_0$ is chosen to be a function of the quark mass. However, in 
the SR calculation, we can obtain a good result of the baryon spectra even if 
$r_0$ is assumed to be constant (see Table~\ref{tab_mass}). 

The confining potential is given by~\cite{ESP1,ESP3} 
\begin{eqnarray}
V_{\rm{CON}}({\vec r}_{ij})
=-C_{\rm{3q}}
-\left\{
\begin{array}{l>{$}l<{$}}
a_c\,(1-e^{-\mu_c r_{ij}}) 
({\vec \lambda}_i^c\cdot{\vec \lambda}_j^c)
& \ \ for NR, \\
a_c\, r_{ij}
({\vec \lambda}_i^c\cdot{\vec \lambda}_j^c)
& \ \ for SR, 
\end{array}
\right. \label{conf1}
\end{eqnarray}
with $a_c$ the effective confinement strength, $\mu_c$ a parameter for the screening effect and 
$C_{3q}$ a parameter for taking account of the vacuum energy. 

To perform the numerical calculation, we use the following identities:
\begin{eqnarray}
\sqrt{p^2+m^2} &=& \sqrt{C}-\frac{1}{\sqrt{\pi}}\int_{0}^{\infty}
\frac{1}{w^2}(e^{-(p^2+m^2)w^2}-e^{-Cw^2})dw , \label{kinetic2} \\
\frac{e^{-ar}}{r} &=& \frac{2}{\sqrt{\pi}}\int_{0}^{\infty}
e^{-\frac{a^2}{4w^2}-r^2w^2}dw , \label{yukawa1}
\end{eqnarray}
where $C$ is an arbitrary, real number. Using these formulas, one can expand the SR kinetic energy and the potentials in terms 
of gaussian functions. 
%
%
%

Varying the variational parameters included in the JAMD wavefunction, the total energy of the system 
\begin{eqnarray}
E^{\pm}=
\frac{\langle \Psi^{\pm}|H|\Psi^{\pm} \rangle}{\langle \Psi^{\pm}|\Psi^{\pm} \rangle}  \label{energy1}
\end{eqnarray}
is minimized. To perform such calculation, it is very convenient to use the frictional cooling method~\cite{ENYO1}, which 
provides the time-development equations for the center, ${\vec Z}_i$, and the width parameter, $\tilde{\nu}_{ij}$, 
as\footnote{
Because the Jacobi coordinates, $({\vec \rho}_i, {\vec X}_i)$, can be transformed into the old set, $({\vec r}_i, {\vec Z}_i)$, 
using the $K$ matrix, it is sufficient to know the time development of the latter variables~\cite{JAMD}. 
The actual calculation is performed under the constraint of ${\vec X}_A = 0$.}
\begin{eqnarray}
\frac{d Z_{i\lambda}}{dt} &=& 
\mu \frac{\partial E^\pm}{\partial Z^*_{i\lambda}}
\ \ \ \ \  (i=1,2,\cdots,A;\ \lambda=x,y,z) , \label{z-eq1} \\
\frac{d\tilde{\nu}_{ij}}{dt} &=&
\mu' \frac{\partial E^\pm}{\partial \tilde{\nu}_{ij}}
\ \ \ \ \  (i>j=1,2,\cdots,A)  , \label{nu-eq1} 
\end{eqnarray}
where $\mu$ and $\mu'$ are arbitrary, negative real numbers.  
After sufficient time steps for the cooling, we can obtain the optimized wavefunction as a function of ${\vec Z}_i$ 
and $\tilde{\nu}_{ij}$.  We refer to this wavefunction as $\Psi^{{\rm JAMD-I}}$.  

In contrast, we consider another wavefunction which is given by superposing the JAMD wavefunction $\Psi$ ($\pm$ is suppressed here) with 
${\vec Z}_i$ calculated in $\Psi^{{\rm JAMD-I}}$ but with ${\tilde{\nu}_{ij}}$ supplied by a geometrical progression~\cite{JAMD} or 
a random number generator~\cite{SVM2}.  In this case, ${\vec Z}_i$ and ${\tilde{\nu}_{ij}}$ are no longer 
the variational parameters. Then, the wavefunction (we call this $\Psi^{{\rm JAMD-II}}$) is given by
\begin{eqnarray}
\Psi^{\rm{JAMD-II}} = c \Psi({\vec Z}_i, \tilde{\nu}_{ij}) + c' \Psi({\vec Z}_i, {\tilde{\nu}'}_{ij}) 
+ c'' \Psi({\vec Z}_i, {\tilde{\nu}''}_{ij}) + \cdots ,  \label{jamd-2}
\end{eqnarray}
where, instead of ${\vec Z}_i$ and ${\tilde{\nu}_{ij}}$, the coefficients, $c, c'', c''', \cdots$, are now variational parameters, 
and they are determined by the Hill-Wheeler equation 
\begin{eqnarray}
\delta(\langle \Psi^{\rm{JAMD-II}}|H|\Psi^{\rm{JAMD-II}} \rangle
-E\langle \Psi^{\rm{JAMD-II}}|\Psi^{\rm{JAMD-II}} \rangle)=0.  \label{hweq}
\end{eqnarray}

\begin{table}
\begin{center}
\caption{\label{tab_para}Values of the parameters.}
\begin{tabular}{llcccccc}
\hline
\multicolumn{1}{l}{}&\multicolumn{1}{l}{}&
\multicolumn{3}{c}{NR~\protect\cite{ESP3}}&\multicolumn{1}{c}{}&\multicolumn{2}{c}{SR~\protect\cite{ESP1}}\\
\multicolumn{1}{c}{Fixed}&\multicolumn{7}{c}{}\\
\cline{1-1}
\multicolumn{1}{l}{$u$, $d$ quark mass}&\multicolumn{1}{l}{$m_u = m_d$\,(MeV)}&
\multicolumn{3}{c}{313}&\multicolumn{1}{c}{}&
\multicolumn{2}{c}{313}\\
\multicolumn{1}{l}{NG bosons}&\multicolumn{1}{l}{$m_{\pi}\,$(fm$^{-1}$)}&
\multicolumn{3}{c}{0.70}&\multicolumn{1}{c}{}&\multicolumn{2}{c}{0.70}\\
\multicolumn{1}{l}{}&\multicolumn{1}{l}{$m_{\eta}\,$(fm$^{-1}$)}&
\multicolumn{3}{c}{2.77}&\multicolumn{1}{c}{}&\multicolumn{2}{c}{2.77}\\
\multicolumn{1}{l}{}&\multicolumn{1}{l}{$m_{K}\,$(fm$^{-1}$)}&
\multicolumn{3}{c}{2.51}&\multicolumn{1}{c}{}&\multicolumn{2}{c}{2.51}\\
\multicolumn{1}{l}{$\sigma$ meson mass}&\multicolumn{1}{l}{$m_{\sigma}\,$(fm$^{-1}$)}&
\multicolumn{3}{c}{3.42}&\multicolumn{1}{c}{}&\multicolumn{2}{c}{3.42}\\
\multicolumn{1}{l}{cutoff}&
\multicolumn{1}{l}{$\Lambda_{\pi}=\Lambda_{\sigma}$\, (fm$^{-1}$)}&
\multicolumn{3}{c}{4.20}&\multicolumn{1}{c}{}&\multicolumn{2}{c}{2.20}\\
\multicolumn{1}{l}{}&
\multicolumn{1}{l}{$\Lambda_{\eta}=\Lambda_{K}$\, (fm$^{-1}$)}&
\multicolumn{3}{c}{5.20}&\multicolumn{1}{c}{}&\multicolumn{2}{c}{2.70}\\
\multicolumn{1}{l}{coupling constant \ }&
\multicolumn{1}{l}{$g^2_{ch}/(4\pi)$\,(fm$^{-1}$)}&
\multicolumn{3}{c}{0.54}&\multicolumn{1}{c}{}&\multicolumn{2}{c}{0.54}\\
\multicolumn{1}{l}{mixing angle}&\multicolumn{1}{l}{$\theta_P\, $($^\circ$)}&
\multicolumn{3}{c}{$-15$}&\multicolumn{1}{c}{}&\multicolumn{2}{c}{$-15$}\\
\multicolumn{1}{l}{confinement}&
\multicolumn{1}{l}{$a_c$\, (MeV)}&\multicolumn{3}{c}{230}&
\multicolumn{1}{c}{}&\multicolumn{2}{c}{110}\\
\multicolumn{1}{l}{}&\multicolumn{1}{l}{$\mu_c$\, (fm$^{-1}$)}&
\multicolumn{3}{c}{0.70}&\multicolumn{1}{c}{}&\multicolumn{2}{c}{---}\\
\multicolumn{1}{l}{OGE}&\multicolumn{1}{l}{$r_0$}&\multicolumn{3}{c}{0.35}&
\multicolumn{1}{c}{}&\multicolumn{2}{c}{0.74}\\
\\
\multicolumn{1}{l}{}&\multicolumn{1}{l}{}&
\multicolumn{3}{c}{NR}&\multicolumn{1}{c}{}&\multicolumn{2}{c}{SR}\\
\multicolumn{1}{l}{}&\multicolumn{1}{l}{}&
\multicolumn{1}{c}{AMD \ }&\multicolumn{1}{c}{JAMD-I \ }&
\multicolumn{1}{c}{JAMD-II}&\multicolumn{1}{l}{}&\multicolumn{1}{c}{JAMD-I \ }&
\multicolumn{1}{c}{JAMD-II}\\
\cline{3-5}\cline{7-8}
\multicolumn{1}{c}{Free}&\multicolumn{7}{l}{}\\
\cline{1-1}
\multicolumn{1}{l}{$s$ quark mass}&\multicolumn{1}{l}{$m_s$\,(MeV)}&
\multicolumn{1}{c}{598}&\multicolumn{1}{c}{587}&
\multicolumn{1}{c}{554}&\multicolumn{1}{c}{}&\multicolumn{1}{c}{562}&
\multicolumn{1}{c}{525}\\
\multicolumn{1}{l}{vacuum}&\multicolumn{1}{l}{$C_{3q}$\, (MeV)}&
\multicolumn{1}{c}{333}&\multicolumn{1}{c}{346}&
\multicolumn{1}{c}{372}&\multicolumn{1}{c}{}&\multicolumn{1}{c}{128}&
\multicolumn{1}{c}{177}\\
\multicolumn{1}{l}{OGE}&\multicolumn{1}{l}{$\alpha_s$}&
\multicolumn{1}{c}{0.858}&\multicolumn{1}{c}{0.761}&
\multicolumn{1}{c}{0.540}&\multicolumn{1}{c}{}&\multicolumn{1}{c}{0.775}&
\multicolumn{1}{c}{0.500}\\
\hline
\end{tabular}
\end{center}
\end{table}
Now we are in a position to show the numerical result for the baryon spectra. 
The parameters in the present calculation are listed in Table~\ref{tab_para}.  In case of the NR calculation, 
we take the parameters given in Ref.~\cite{ESP3}. We assume that 
the value of $r_0$ in $V_{\rm OGE}$ is flavor-independent. In the SR calculation, referring to Ref.~\cite{ESP1}, 
we determine the parameters in the potentials.  Finally, three parameters, $C_{3q}$, $\alpha_s$ and $m_s$, remain. 
Then, the nucleon ($N$) mass is reproduced by tuning $C_{3q}$, while 
$\alpha_s$ is chosen so as to fit the $N$-$\Delta$ mass difference. The strange-quark mass, $m_s$, is determined from fit to 
the $N$-$\Omega$ mass difference. 

\begin{table}
\begin{center}
\caption{\label{tab_mass}Calculated baryon masses (in MeV). The excited baryons are calculated in the JAMD-II.}
\begin{tabular}{lcccccccccc}
\hline
State& &\multicolumn{4}{c}{NR}&\multicolumn{1}{c}{}
&\multicolumn{2}{c}{SR}&\multicolumn{1}{c}{}&Experiment\\
& &AMD \ &JAMD-I \ &JAMD-II \ & Faddeev\cite{ESP3} & & JAMD-I \ & JAMD-II & & \\
\cline{1-1}\cline{3-6}\cline{8-9}\cline{11-11}
$N(1/2^+)$& & 939 & 939 & 939  & 939 &  & 939 & 939& & 939 \\
$N(1/2^-)$ && 1481 & 1460 & 1409  & 1411& & 1522 &1480 &&1553\\
$N^*(1/2^+)$ && --- & --- & 1423  & 1435 & &--- &1409& &1440\\
$\Delta(3/2^+)$ && 1230 & 1232 & 1234  & 1232 & &1235 & 1236& &1232\\
$\Delta^*(3/2^+)$ && --- & --- & 1589  & ---&  &--- & 1601&&1600\\
\cline{1-1}\cline{3-6}\cline{8-9}\cline{11-11}
$\Sigma(1/2^+)$   && 1315 & 1264 & 1246  & 1213 & & 1244 & 1214& &1193\\
$\Sigma^*(1/2^+)$  && --- & 1400 & 1678  & 1644 & & --- &1553& &1660\\
$\Sigma(3/2^+)$ && 1435 & 1404 & 1396  & 1398 & &1384 &1382& & 1385\\
$\Sigma(1/2^-)$ && 1722 & 1639 & 1631  & 1598  &&1726 &1678 &&1620\\
$\Lambda(1/2^+)$ && 1166 & 1160 & 1135  & 1122  &&1139 &1120 &&1116\\
$\Xi(1/2^+)$  && 1418 & 1401 & 1375  & 1351  &&1374 &1359 &&1318\\
$\Omega(3/2^+)$  && 1673 & 1673 & 1673  & 1650  &&1671 &1670 &&1672\\
\hline
\end{tabular}
\end{center}
\end{table}

%
\begin{figure}
\begin{center}
\includegraphics[scale=1.0]{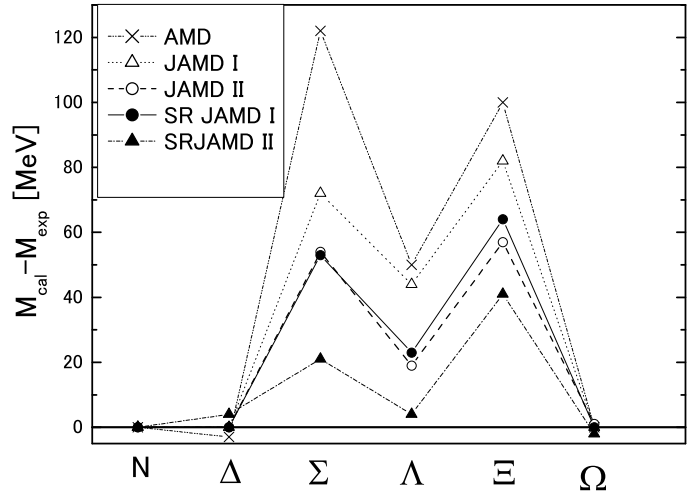}
\caption{\label{fig_mass} Difference between the calculated mass and the experimental value.}
\end{center}
\end{figure}
\begin{figure}
\begin{center}
\includegraphics[scale=1.0]{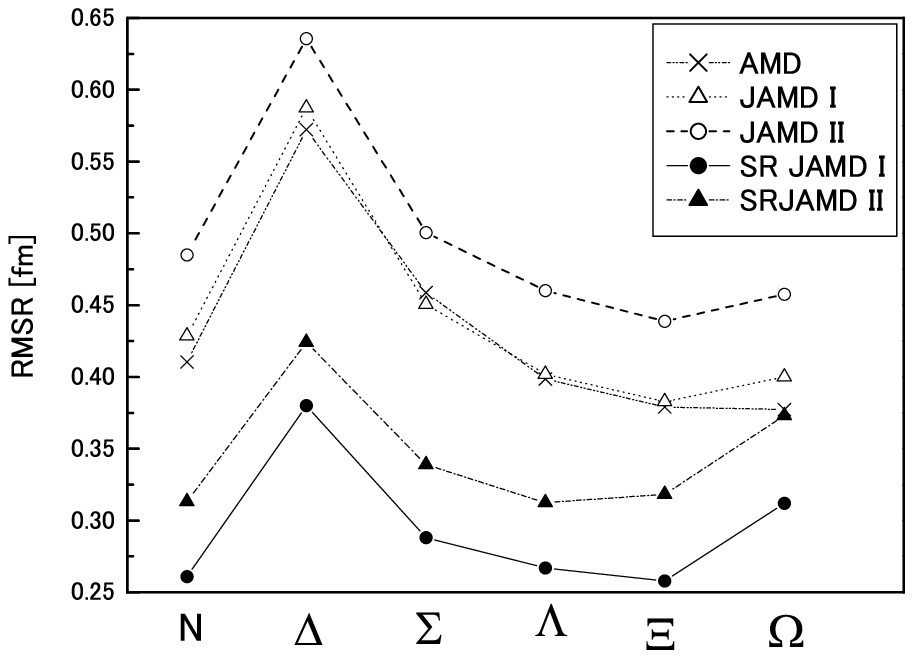}
\caption{\label{fig_rmsr} Root-mean-square radius.}
\end{center}
\end{figure}

In Table~\ref{tab_mass}, we present the results of AMD and JAMD, and compare them with the 
Faddeev calculation~\cite{ESP3}. Furthermore, in Fig.~\ref{fig_mass}, the deviation of the calculated, ground-state mass from 
the observed one is shown to illustrate the model dependence of the baryon masses.  The present result clearly show that 
the JAMD approach is much better than AMD in describing the baryon spectra. This fact 
implies the importance of the quark-quark correlation in the baryon structure. In particular, within the NR calculation, 
the JAMD-II result is very close to the spectra given by the Faddeev approach. 

In the SR calculation, the present approach again reproduces the spectra of the low-lying light and strange baryons very well.  
In particular, the JAMD-II can provides the precise result as in the Faddeev calculation. 
It is also remarkable that 
the level ordering of the lowest 
positive- and negative-parity states in the nucleon spectra can be correctly reproduced in the SR calculation~\cite{SVM1,ESP1}. 
This fact certainly results from the relativistic kinematics, because, in the NR calculation, we cannot produce the correct level ordering. 
It is, however, necessary to study further to obtain a quantitative result of $N(1/2^-)$ and $N^*(1/2^+)$. 
%

In Fig \ref{fig_rmsr}, we show the root-mean-square (rms) radius of the baryon calculated using the quark wavefunction. 
It is noticeable that the rms radius is very small in the SR calculation, whereas, in the NR calculation, it has the reasonable size. 
It should, however, be noticed that the meson cloud surrounding the baryon core contributes to the present 
value additionally. The tendency of the rms radii of the low-lying baryons does not depend much on the choice of the model. 

In summary, we have proposed a new approach of molecular dynamics based on AMD, in which the many-body correlation can be 
considered correctly. In this paper, we applied it to the spectra of the low-lying, light and strange baryons.  
It is shown that the inclusion of the quark-quark correlation 
is very vital to predict the precise spectra, and that the relativistic effect is also important to correct the level ordering. 
The JAMD approach can reproduce the baryon spectra very well. 
In particular, the spectra given by the JAMD-II calculation are very similar to the Faddeev result.  
Although we study the ordinary baryons in this paper, the present approach is very promising even in the calculation of a system containing 
four, five or more quarks.  Thus, it is very intriguing to calculate the spectra of exotic hadrons like pentaquark, 
tetraquark, etc~\cite{exotic}.  
%

%
\begin{acknowledgements}
T.W. and K.S. thank A. Valcarce for valuable discussions on the Faddeev result in the baryon spectra. 
This work was supported by Academic Frontier Project (Holcs, Tokyo University of Science, 2005) of MEXT.
\end{acknowledgements}
%

%


\clearpage
\begin{thebibliography}{99}
%
\bibitem{lattice} For a review, Proc. of The XXV International Symposium on Lattice Field Theory, PoS(LATTICE 2007). 
%
\bibitem{saito} K. Saito, K. Tsushima, A.W. Thomas, Prog. Part. Nucl. Phys. 58 (2007) 1.   
\bibitem{cqm}A. de R{\'u}jula, H. Georgi, S. L. Glashow, Phys. Rev. D 12 (1975) 147; \\
N. Isgur, G. Karl, Phys. Rev. D 18 (1978) 4187; \\
M. Oka, K. Yazaki, Phys. Lett. B90 (1980) 41; Prog. Theor. Phys. 66 (1981) 556.   
%
\bibitem{tetra} For example, S.-K. Choi et al., Belle Collaboration, Phys. Rev. Lett. 93 (2003) 26200; \\
See also the homepage of the Particle Data Group, http://pdg.lbl.gov/. 
%
\bibitem{SVM1}L. Ya. Golzman, W. Plessas, K. Varga, R. F. Wagenbrunn, Phys. Rev. D 58 (1998) 094030.
%
\bibitem{ESP1}H. Garcilazo, A. Valcarce, Phys. Rev. C 68
 (2003) 035207.
%
%
\bibitem{ESP3}A. Valcarce, H. Garcilazo, J. Vijande, Phys. Rev. C 72
 (2005) 025206.
%
\bibitem{ESP4}A. Valcarce, H. Garcilazo, F. Fern{\'a}ndez, P. Gonz{\'a}lez, Rep. Prog. Phys. 68 (2005) 965.
%
\bibitem{fmd} H. Feldmeier, Nucl. Phys. A 515 (1990) 147; \\
H. Feldmeier, J. Schnack, Rev. Mod. Phys. 72 (2000) 655.
%
\bibitem{ONO} H. Horiuchi, Nucl. Phys. A 522 (1991) 257c; \\
A. Ono, H. Horiuchi, T. Maruyama, A. Ohnishi, Phys. Rev. Lett. 68 (1992) 2898.
%
\bibitem{ENYO1} Y. Kanada-En'yo, H. Horiuchi, A. Ono, Phys. Rev. C 52 (1995) 628.
%
%
\bibitem{ENYO21}Y. Kanada-En'yo, H. Horiuchi, Prog. Theor. Phys. Suppl. 142 (2001) 205.
%
\bibitem{JAMD}T. Watanabe, S.Oryu, 
Prog. Theor. Phys. 116 (2006) 429.
%
\bibitem{SVM2}Y. Suzuki, K. Varga, ${\it Stochastic Variational Approach to Quantum Mechanical Few-Body Problems}$
 (Springer-Verlag, Berlin, 1997).
%
\bibitem{CRCGV1}M. Kamimura, Phys. Rev. A 38 (1988) 621.
%
\bibitem{watanabe} T. Watanabe, ``A few-body system in AMD", Doctor thesis (2005), unpublished.
%
%
\bibitem{exotic}Y. Kanada-En'yo, O. Morimatsu, T. Nishikawa, 
Phys. Rev. C 71 (2005) 045202; \\
%
Y. Kanada-En'yo, O. Morimatsu, T. Nishikawa, 
Phys. Rev. D 71 (2005) 094005.
%

%
%
%
%
%
%
\end{thebibliography}
\end{document}